\documentclass{article}
\usepackage{graphicx}
\usepackage{amsmath}
\usepackage{amssymb}
\usepackage{booktabs}
\usepackage[numbers,sort&compress]{natbib}
\usepackage{hyperref}
\usepackage{tikz}
\usetikzlibrary{positioning, arrows.meta, shapes.geometric}
\usepackage{subcaption}
\usepackage{authblk}

\title{Serial Dependence Emerges from Optimizing Vision for Noise Robustness}
\author[1]{Hiroshi Higashi}
\author[2]{Keigo Sugihara}
\author[2]{Yuichi Tanaka}
\affil[1]{Faculty of Engineering Science, Kansai University}
\affil[2]{Graduate School of Engineering, The University of Osaka}
\date{\today}

\begin{document}

\maketitle

\begin{abstract}
Human visual perception is dynamically shaped by the recent past, a phenomenon known as serial dependence.
Mechanistic models show that a residual memory trace can generate this dependency, yet its functional purpose (why the visual system should carry information forward in time at all) remains debated.
We tested the hypothesis that serial dependence emerges as a normative consequence of optimizing vision for robustness against sensory noise.
We trained recurrent autoencoders (a CNN encoder, a ConvGRU memory module, and a CNN decoder) to reconstruct natural videos corrupted by Gaussian noise, with no objective referring to perceptual history.
Models that used their temporal memory achieved lower reconstruction error than memory-free counterparts, and this advantage widened as training noise increased, indicating that integrating past context is an adaptive denoising strategy.
Probing the trained models with Gabor and dot sequences, we found that memory gave rise to serial dependence spontaneously: an attractive bias toward previous orientations and a spatial ``continuity field'' toward recently seen locations, both strengthening with training noise.
Across temporal delays, the model's serial dependence was strongest for the immediately preceding stimulus and decayed monotonically.
This profile quantitatively matched human orientation judgments, including an independent public dataset, once a known response-autocorrelation confound was controlled: an apparent non-monotonicity in one dataset reflected this measurement artifact rather than a genuine model--human mismatch.
These findings suggest that serial dependence need not be an idiosyncratic quirk but can arise as a byproduct of optimizing for noise robustness, offering a normative account of perceptual history-dependence that complements existing mechanistic explanations and links behavior to an underlying computational objective.
\end{abstract}

\section{Introduction}
Visual perception is a dynamic process continuously influenced by recent sensory history.
Perceptual judgments are pulled toward recently viewed stimuli, a phenomenon termed serial dependence \citep{Fischer2014-rv, Burr2014-uo}.
This attractive bias is ubiquitous, influencing the perception of features such as orientation \citep{Fischer2014-rv, Fritsche2017-yn, Murai2021-sd, Alais2017-tn}, spatial position \citep{Manassi2018-be}, numerosity \citep{Cicchini2014-sg}, proportion \citep{Hashiguchi2026}, face identity \citep{Liberman2014-kp}, and facial expression \citep{Liberman2018-tb}.

Recently, we proposed SDNet, a mechanistic model that explains serial dependence as an emergent property of evidence accumulation \citep{Higashi2025-sd}.
By pairing a convolutional neural network (CNN) with a recurrent neural network (RNN) that retains a residual memory trace of past decisions, this model reproduced human-like serial dependence.
That study, however, characterized serial dependence at the algorithmic level, detailing \textit{how} a memory trace can generate the bias, while leaving open the question at the level of computational theory \citep{Marr1982}: \textit{why} the visual system should adopt a memory-dependent strategy at all, and what problem such a strategy solves.
Optimality-based theories offer a candidate answer, proposing that serial dependence enhances perceptual stability in noisy environments \citep{Manassi2024-ob, Cicchini2024-jp}, with the recent past acting as a prior that refines the representation of current, ambiguous sensory inputs \citep{Cicchini2024-jp, Stocker2006-nz, Ernst2002-hu}.
Yet these accounts have largely been formulated as abstract inference rules, leaving unresolved whether the same behavior arises spontaneously in a system that is simply optimized to perform under noise.

Here we bridge the mechanistic and normative levels of analysis by asking whether serial dependence emerges as an optimal solution to a concrete computational problem: recovering a stable percept from noisy, time-varying input.
We adopt a task-optimized modeling strategy, training artificial networks on an ecologically motivated objective and examining the perceptual regularities that emerge, an approach that has yielded functional explanations of neural and perceptual phenomena \citep{Kriegeskorte2018-ir}.
Concretely, we trained recurrent autoencoders to reconstruct natural video frames from noisy inputs and asked whether, and how strongly, serial dependence emerges as a function of environmental uncertainty.
Because this hypothesis concerns the objective rather than any specific circuit, it predicts that serial dependence should be a generic consequence of optimizing for temporal denoising in any system, biological or artificial, that contends with sensory noise.
Our results thereby provide a functional justification for the residual memory traces posited by mechanistic accounts \citep{Higashi2025-sd, Tang2018-uj}, grounding them in a normative principle.
This framing sets the present work apart from our mechanistic SDNet, which generated serial dependence through a recurrent memory of past \textit{decisions} embedded in a discrimination task \citep{Higashi2025-sd}.
The training objective of the present networks makes no reference to perceptual history (they are optimized purely to reconstruct clean frames), so serial dependence emerges as an unintended byproduct rather than through history-dependent machinery.
Moving from a mechanistic account (\textit{how}) to a normative one (\textit{why}) also yields quantitative predictions that were untested in prior work: that the bias scales with sensory uncertainty, and that the same objective gives rise to a spatial continuity field.
Comparing the model directly against human orientation judgments, including an independent public dataset, we find that human serial dependence is monotonic and quantitatively matches the model once a known response-autocorrelation confound is controlled.

\section{Results}

\begin{figure}[tp]
\centering
\includegraphics[width=\textwidth]{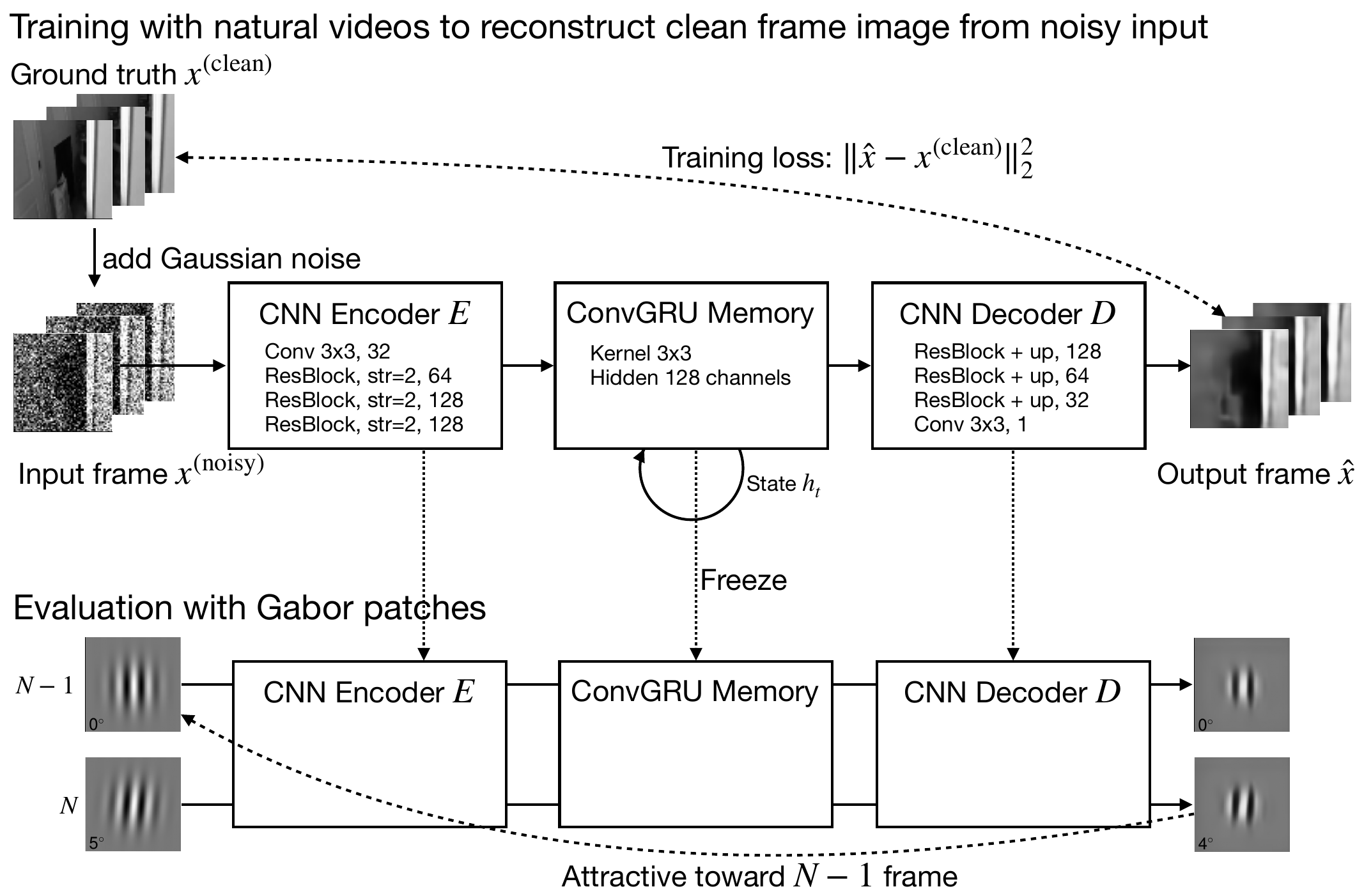}
\caption{
Overview of the model and the experimental paradigm.
  \textbf{(Top) Training.} A recurrent autoencoder (a CNN encoder $E$, a ConvGRU memory module that maintains a temporal state $h_t$, and a CNN decoder $D$) is trained on natural video sequences (Charades-Ego) to reconstruct each clean frame $x^{(\mathrm{clean})}_t$ (ground truth) from an input $x^{(\mathrm{noisy})}_t$ corrupted by additive Gaussian noise. The network is optimized end-to-end to minimize the reconstruction loss $\|\hat{x}_t - x^{(\mathrm{clean})}_t\|_2^2$ between its output $\hat{x}_t$ and the clean ground truth, with no objective referring to perceptual history. The encoder $E$ (Conv $3\times3$; three stride-2 ResBlocks, 64/128/128 channels), the ConvGRU memory ($3\times3$ kernel, 128 hidden channels; recurrent state $h_t$), and the decoder $D$ (three ResBlock+upsample stages, 128/64/32 channels; final Conv $3\times3$) together form the recurrent pathway.
\textbf{(Bottom) Evaluation.} The trained weights are then frozen and the model is probed with sequences of Gabor patches. Because the memory carries information forward from the preceding frame ($N-1$), the reconstruction of the current frame ($N$) is attracted toward the previous orientation: here a $5^\circ$ stimulus preceded by a $0^\circ$ stimulus is reconstructed at $4^\circ$, revealing an emergent attractive serial dependence.
}
\label{fig:architecture}
\end{figure}

We trained recurrent autoencoders to reconstruct clean frames from noise-corrupted natural videos, then froze the learned weights and probed each model with controlled Gabor and dot sequences to characterize the perceptual biases that emerged (Figure \ref{fig:architecture}).

\subsection{Functional benefits of memory}
\begin{figure}[tp]
\centering
\includegraphics[width=\textwidth]{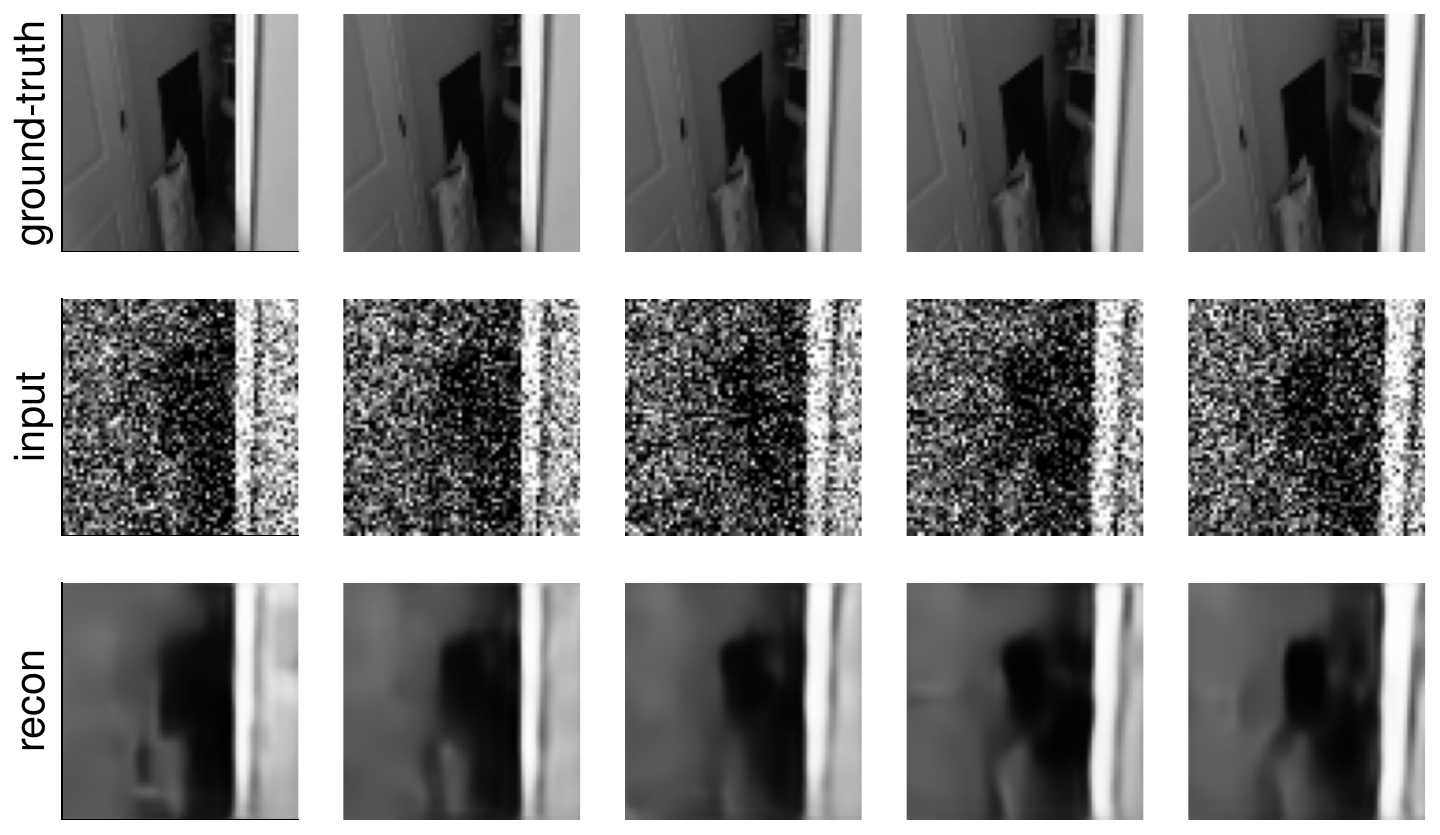}
\caption{
  Video reconstruction by the model trained with $\sigma=0.3$.
  Columns show successive frames from $N-4$ to $N$.
  Rows: ground-truth clean frames $x^{(\mathrm{clean})}_t$, noisy input $x^{(\mathrm{noisy})}_t$ (clean frame corrupted by additive Gaussian noise, $x^{(\mathrm{noisy})}_t = \mathrm{clip}(x^{(\mathrm{clean})}_t + \epsilon_t,\, 0,\, 1)$), and reconstructed output $\hat{x}_t$.
}
\label{fig:vis_video}
\end{figure}

To verify the utility of temporal integration, we compared the reconstruction error (mean squared error, MSE) of models with and without memory.
Qualitatively, the recurrent model removed Gaussian noise while preserving structural features (Figure \ref{fig:vis_video}).

\begin{figure}[tp]
  \centering
  \includegraphics[width=0.48\textwidth]{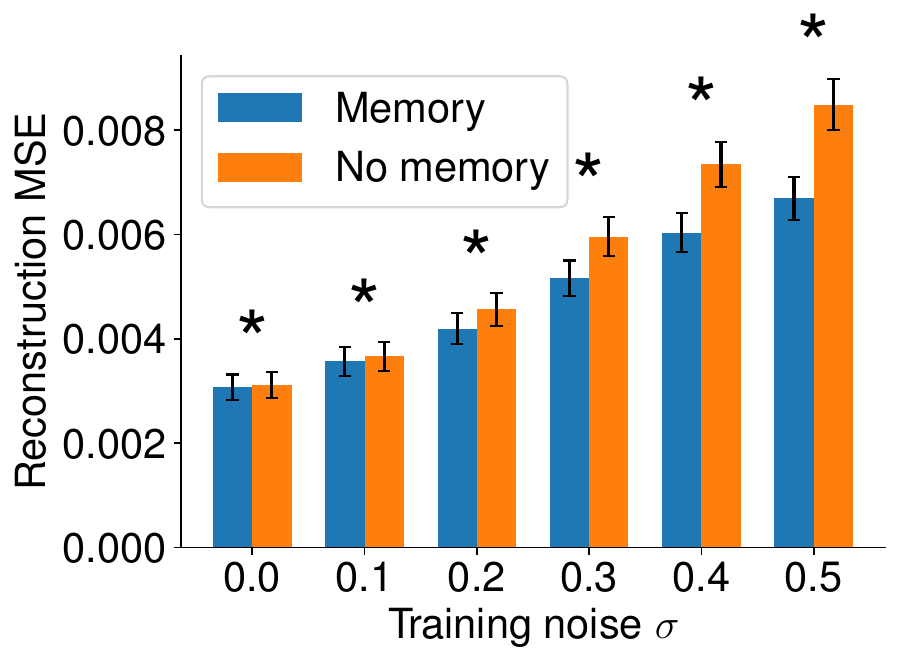}
  \caption{
    Comparison of reconstruction error (MSE).
    Asterisks indicate significant differences between the Memory and No-memory models at each noise level ($p < 0.05$, paired $t$-test).
    }
    \label{fig:mse_comparison}
\end{figure}

Quantitatively, the model using memory (Memory) outperformed the memory-reset model (No-memory) at all noise levels (Figure \ref{fig:mse_comparison}).
We evaluated 100 randomly sampled 5-frame video sequences.
For the Memory condition, the full sequence was processed recurrently and the MSE was computed on the last reconstructed frame.
For the No-memory condition, only the last frame was fed to the same trained model with the hidden state reset to zero, so that no temporal context was available.
At $\sigma=0.3$, the Memory model achieved a significantly lower mean MSE than the No-memory model (0.0052 vs.\ 0.0060; $t(99) = -14.4, p < 0.001$, paired $t$-test).
This performance advantage increased monotonically with noise, from a small difference at $\sigma=0.0$ to a gap of roughly $0.0018$ at $\sigma=0.5$, indicating that temporal integration is most valuable when noise is high.

\subsection{Emergence of serial dependence under noise}
\begin{figure}[tp]
    \centering
    \begin{minipage}{0.48\textwidth}
        \centering
        \includegraphics[width=\textwidth]{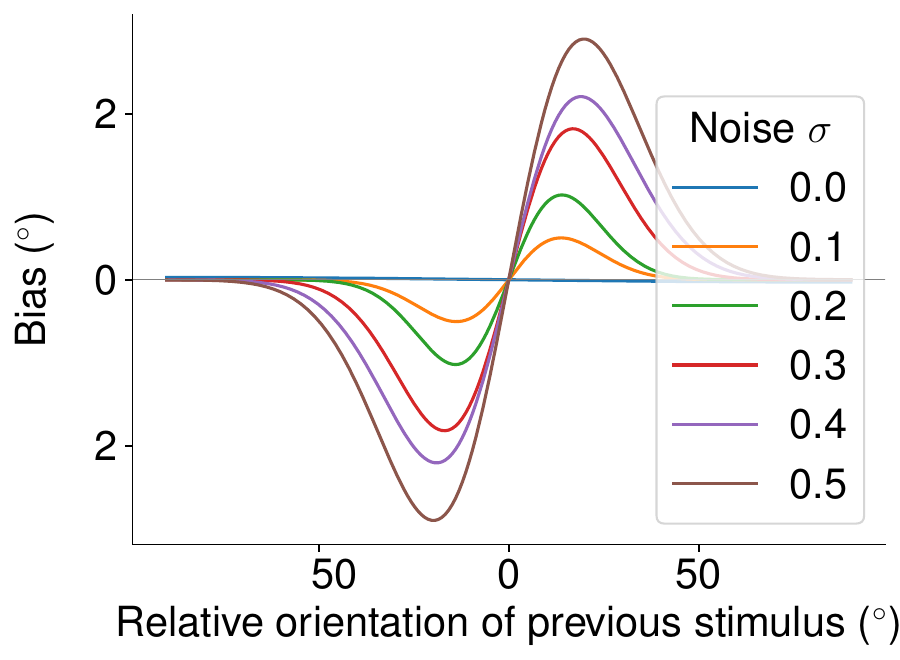}
        \subcaption{}
        \label{fig:sd_curves}
    \end{minipage}
    \\
    \begin{minipage}{0.48\textwidth}
        \centering
        \includegraphics[width=\textwidth]{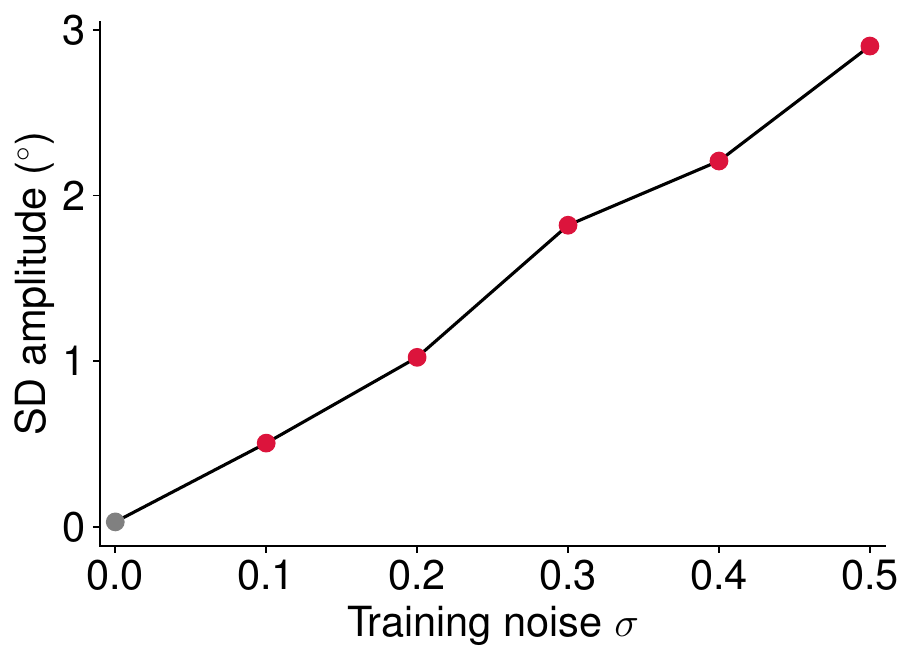}
        \subcaption{}
        \label{fig:sd_strength}
    \end{minipage}
    \hfill
    \begin{minipage}{0.48\textwidth}
        \centering
        \includegraphics[width=\textwidth]{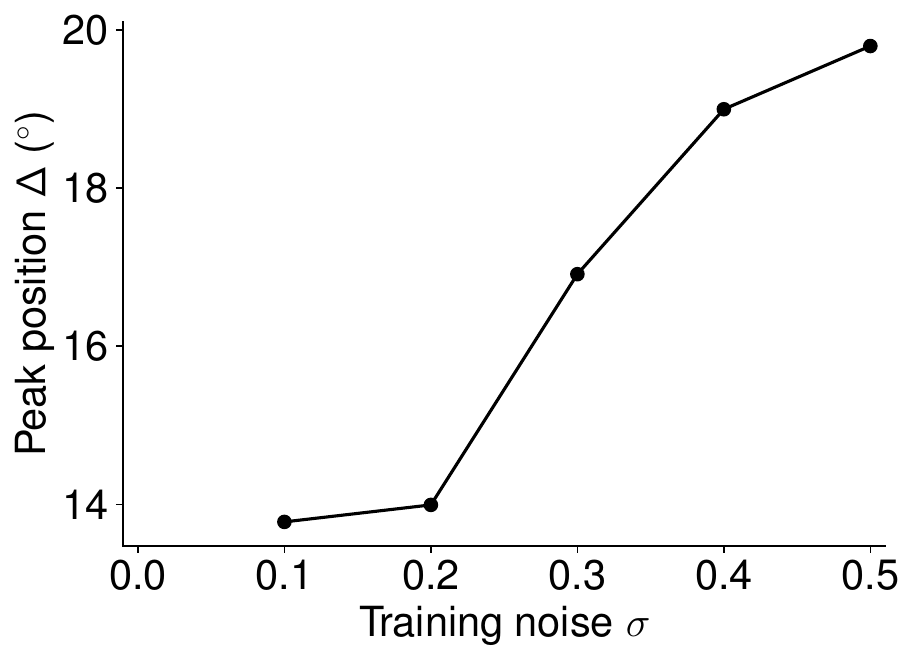}
        \subcaption{}
        \label{fig:sd_delta}
    \end{minipage}
    \caption{
      Emergent serial dependence.
      (a) Adjustment errors as a function of the relative orientation between the current and preceding stimuli for models trained at varying noise levels.
      Dots represent individual model means, and solid lines represent the fitted DoG curves.
      (b) SD amplitude as a function of training noise.
      Red dots denote SD amplitudes significantly greater than chance ($p < 0.05$, permutation test).
      (c) SD peak position (Delta) as a function of training noise. Only models with significant serial dependence are plotted.
    }
    \label{fig:sd_results}
\end{figure}

We next examined whether temporal integration yields serial dependence (Figure \ref{fig:sd_results}).
To probe orientation serial dependence, we presented each trained model with five-frame sequences of Gabor stimuli at random orientations and read out the perceived orientation of the final frame by matching its reconstruction to a bank of Gabor templates; the adjustment error was defined as the signed difference between this estimate and the true orientation.
Serial dependence was then quantified as the peak amplitude of a Derivative-of-Gaussian (DoG) curve fitted to the adjustment error as a function of the relative orientation between the current and preceding stimulus, with a positive amplitude denoting an attractive bias toward the previous stimulus (Materials and methods).
As training noise increased, perceptual estimates were pulled more strongly toward the preceding stimulus (Figure \ref{fig:sd_curves}).
Permutation tests confirmed significant serial dependence ($\text{SD amplitude} > 0$) in every model trained with noise ($\sigma \ge 0.1$, all $p \le 0.002$), whereas the noise-free model showed no significant bias ($\sigma = 0.0$: amplitude $0.03^\circ$, $p = 0.76$).
The SD amplitude increased monotonically with environmental uncertainty, from $0.51^\circ$ at $\sigma=0.1$ to $2.90^\circ$ at $\sigma=0.5$ (Figure \ref{fig:sd_strength}).
Across all models with significant serial dependence, the DoG curves peaked at small orientation differences (peak position $\Delta \approx 14$--$20^\circ$; Figure \ref{fig:sd_delta}), the characteristic tuning of orientation serial dependence.

\begin{figure}[tp]
\centering
\includegraphics[width=\textwidth]{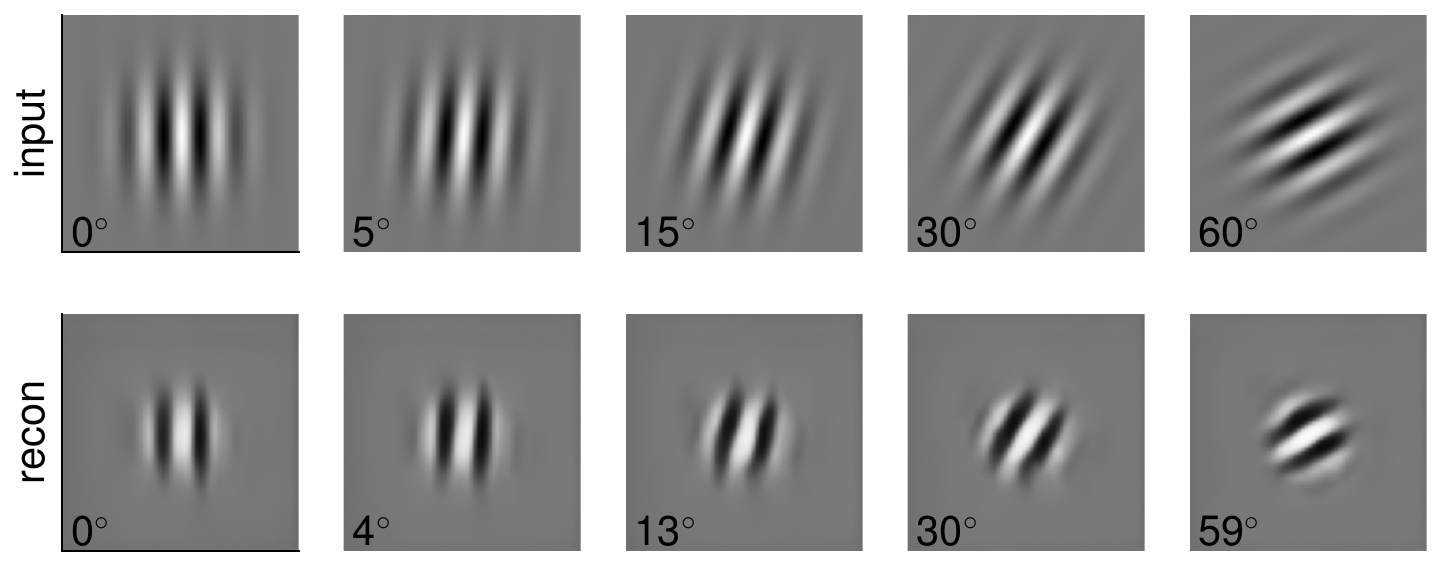}
\caption{
  Reconstruction of Gabor stimuli by the model trained with $\sigma=0.3$.
  Columns show successive frames from $N-4$ to $N$.
  Top row: Ground truth stimuli.
  Bottom row: Reconstructed outputs.
  True and estimated orientations are indicated at the bottom.
}
\label{fig:vis_gabor}
\end{figure}

Figure~\ref{fig:vis_gabor} shows reconstructed Gabor sequences for the $\sigma=0.3$ model.
When the orientation difference between successive frames was small, reconstructed images exhibited an attractive bias toward the preceding frame.
Conversely, when the orientation difference was large, reconstructions were less influenced by the prior frame, consistent with the characteristic tuning of serial dependence.

\begin{figure}[tp]
    \begin{minipage}{0.48\textwidth}
        \centering
        \includegraphics[width=\textwidth]{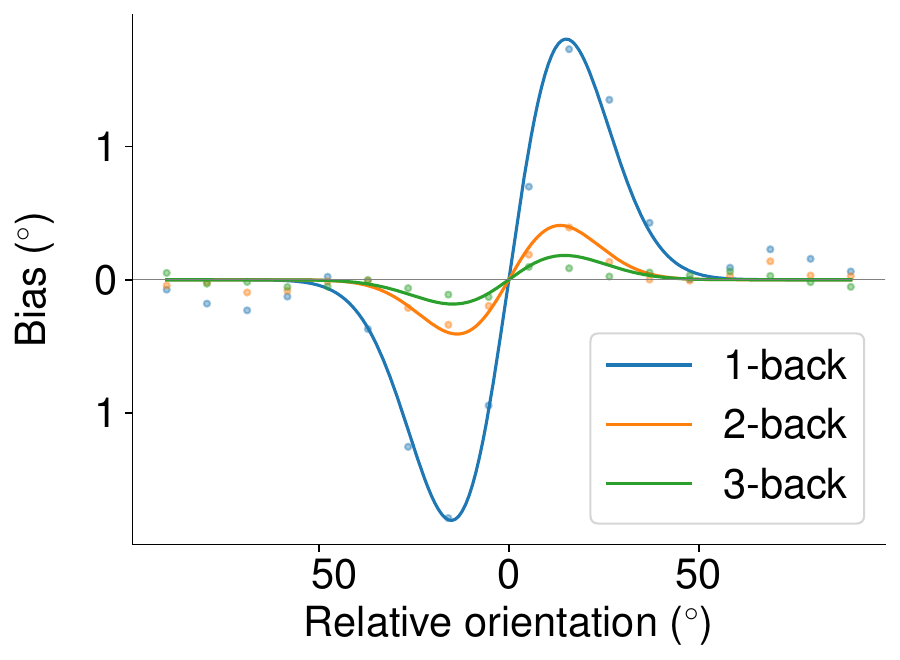}
        \subcaption{}
    \end{minipage}
    \hfill
    \begin{minipage}{0.48\textwidth}
        \centering
        \includegraphics[width=\textwidth]{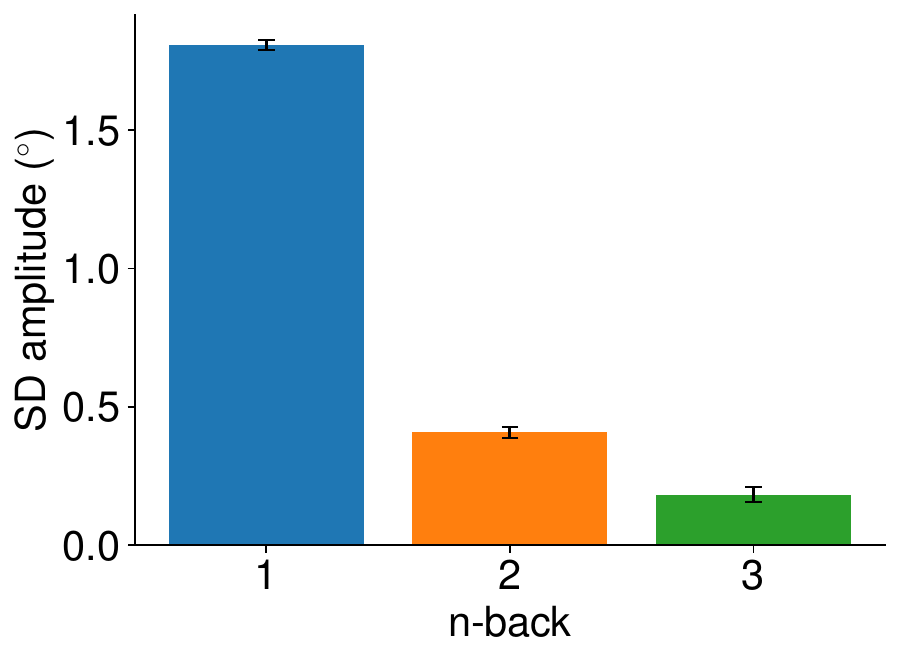}
        \subcaption{}
    \end{minipage}
    \caption{
      Temporal decay of serial dependence in model predictions ($N=100$ models trained with $\sigma=0.3$), computed with the response-autocorrelation-corrected pipeline later applied to the human data.
      Left: combined adjustment error curves as a function of relative orientation between the current and 1-, 2-, and 3-back stimuli. Dots represent binned model means; solid lines are the group-level DoG fits.
      Right: peak SD amplitude across $n$-back conditions (mean $\pm$ SEM), showing a monotonic temporal decay.
    }
    \label{fig:nback_curves}
\end{figure}

To assess robustness, we trained 100 independent models with $\sigma=0.3$, each on a different randomly shuffled subset of the training videos.
All models exhibited attractive biases (Figure \ref{fig:nback_curves}; mean 1-back amplitude $1.807 \pm 0.018^\circ$, $p<0.001$), indicating that serial dependence emerges reliably from optimizing for noise robustness.

Significant serial dependence was also present for 2-back (mean amplitude $0.408 \pm 0.020^\circ$, $p<0.001$) and 3-back stimuli (mean amplitude $0.183 \pm 0.027^\circ$, $p<0.001$; Figure \ref{fig:nback_curves}). The SD amplitude decayed monotonically over temporal distance, falling to less than a quarter of its 1-back value at 2-back. We compare this temporal profile directly with human behavior below.

\subsection{Human orientation serial dependence is monotonic and matches the model}
We next asked whether the emergent serial dependence quantitatively mirrors human perception, comparing the 100 models ($\sigma=0.3$) against human orientation-adjustment data in which observers reproduced the orientation of a briefly presented grating.
To place model and observers on a common footing, we treated each model's per-frame reconstruction as a ``response'' and passed it through the identical trial-level analysis pipeline used for the human data (Materials and methods).

An initial comparison against our in-house dataset ($N=83$) suggested a qualitative mismatch: whereas the model's serial dependence decayed monotonically with lag, the raw in-house profile was non-monotonic, its 2-back amplitude ($3.81^\circ$) exceeding the 1-back amplitude ($2.73^\circ$).
A peak at 2-back is atypical for orientation serial dependence and can arise spuriously from autocorrelation in successive adjustment responses rather than from a genuine perceptual dependence on older stimuli \citep{Esposito2026-ac}.
We therefore reanalyzed both model and human data after regressing out the influence of the 1-back \textit{response} before estimating the dependence on the 2- and 3-back \textit{stimuli} \citep{Esposito2026-ac} (Materials and methods).
Under this correction the in-house 2-back amplitude fell markedly ($3.81^\circ \rightarrow 2.12^\circ$) and the profile became monotonic (Table \ref{tab:human_model_comparison}), indicating that much of the apparent 2-back peak was a measurement artifact; a residual 2-back attraction nonetheless remained significant in this dataset ($2.12^\circ$, $p=0.036$).

\begin{figure}[tp]
    \begin{minipage}{0.48\textwidth}
        \centering
        \includegraphics[width=\textwidth]{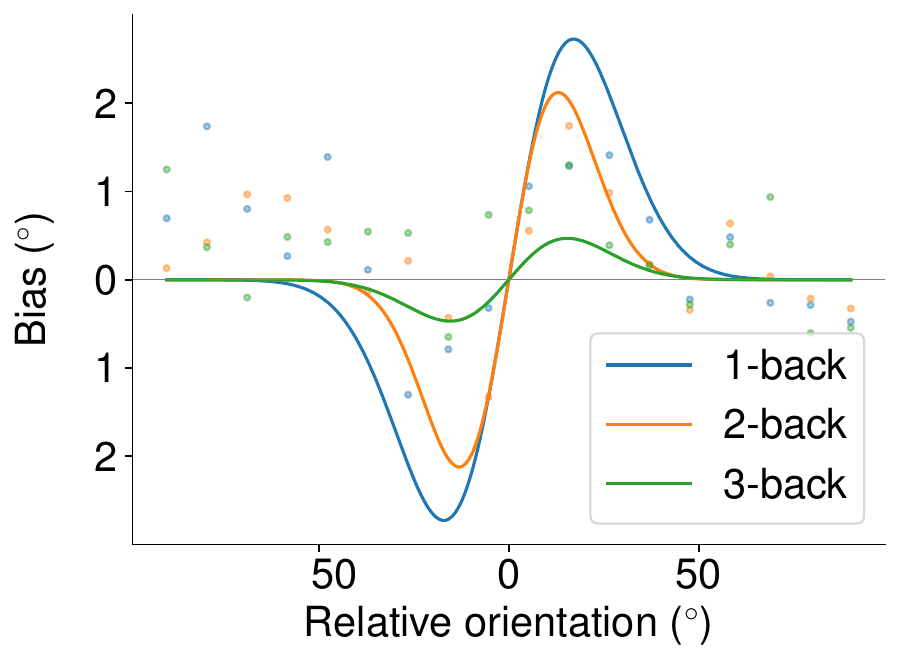}
        \subcaption{}
    \end{minipage}
    \hfill
    \begin{minipage}{0.48\textwidth}
        \centering
        \includegraphics[width=\textwidth]{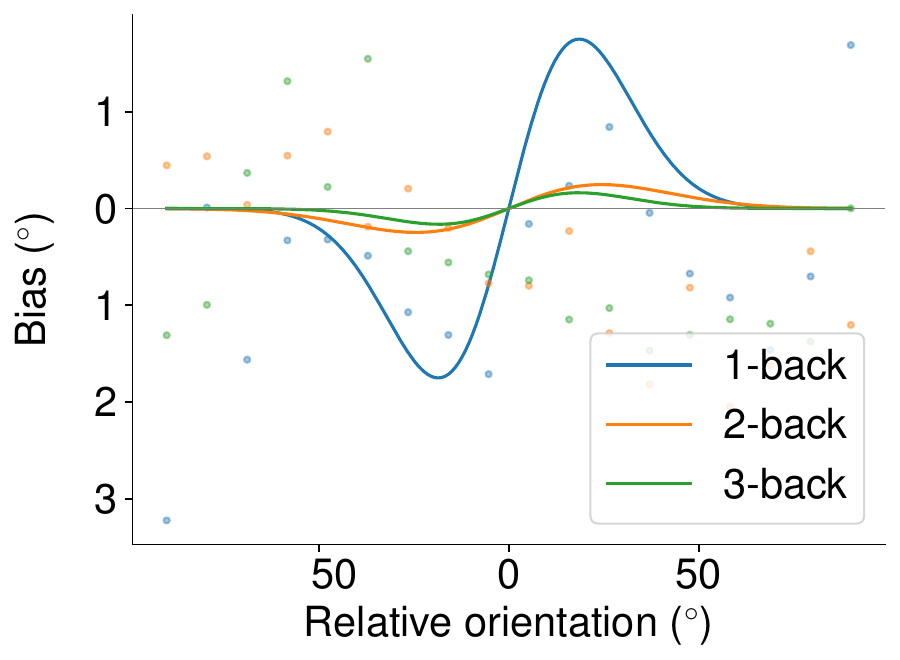}
        \subcaption{}
    \end{minipage}
    \caption{
      Response-autocorrelation-corrected serial dependence in human orientation judgments.
      Combined adjustment-error curves as a function of the relative orientation between the current and 1-, 2-, and 3-back stimuli; dots are binned participant means and solid lines the group-level DoG fits.
      (a) In-house dataset ($N=83$): a dominant 1-back attraction with a smaller residual 2-back component.
      (b) Independent public dataset (Pascucci et al.\ 2019, $N=20$): a monotonic, 1-back-dominant profile.
    }
    \label{fig:human_nback_curves}
\end{figure}

To test whether this corrected, monotonic profile generalizes beyond our own data, we analyzed an independent, publicly available orientation-adjustment dataset \citep{Pascucci2019-tj} ($N=20$, pooled across two experiments) with the identical pipeline.
This dataset showed no 2-back peak even before correction, and after correction its serial dependence was strongly 1-back-dominant and monotonic (1-back $1.75^\circ$, $p<0.001$; 2-back $0.25^\circ$, $p=0.34$; 3-back $0.16^\circ$, $p=0.39$; Table \ref{tab:human_model_comparison}, Figure \ref{fig:human_nback_curves}).

\begin{figure}[tp]
  \centering
  \includegraphics[width=0.6\textwidth]{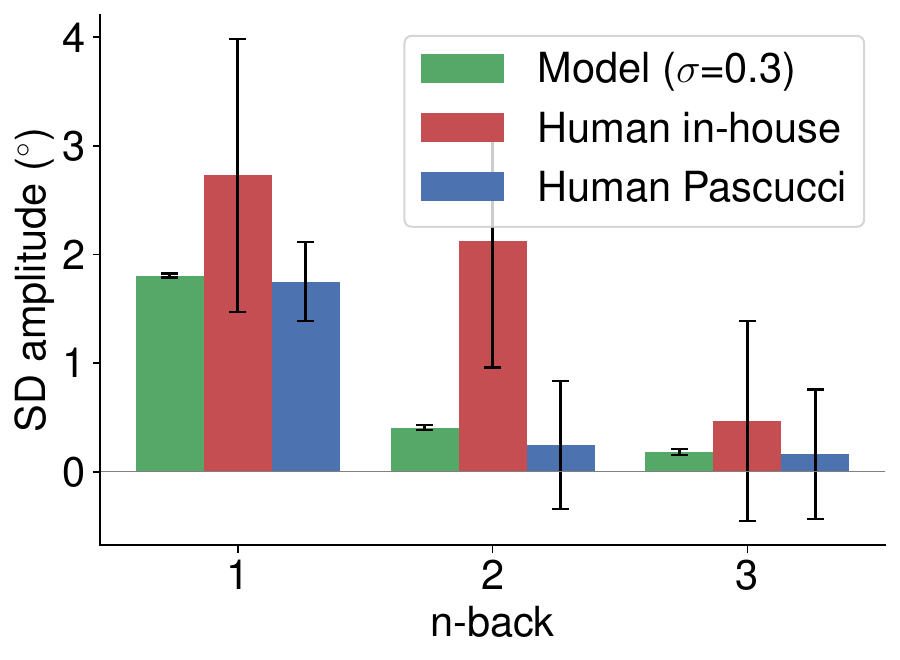}
  \caption{
    Model--human comparison of corrected $n$-back serial dependence.
    Peak SD amplitude (mean $\pm$ SEM) for the model ($\sigma=0.3$, $N=100$), the in-house human data ($N=83$), and the independent public dataset (Pascucci et al.\ 2019, $N=20$), all computed with the identical response-autocorrelation-corrected pipeline.
    Model and public data are strongly 1-back-dominant, decay monotonically, and are statistically indistinguishable at 1-back; the in-house data show a larger residual 2-back component (see text).
  }
  \label{fig:model_vs_human}
\end{figure}

This corrected human profile closely matched the model (Figure \ref{fig:model_vs_human}).
The public-data 1-back amplitude ($1.75^\circ$) was indistinguishable from the model's ($1.81^\circ$), and both decayed monotonically to near-zero by 3-back ($0.16^\circ$ vs.\ $0.18^\circ$).
Thus, once the response-autocorrelation confound is controlled, human orientation serial dependence is monotonic and quantitatively consistent with a system optimized purely for noise-robust reconstruction: most cleanly in the independent public dataset, and in the same direction in our in-house data.

\begin{table}[t]
\centering
\caption{Response-autocorrelation-corrected serial dependence DoG amplitudes (mean $\pm$ SEM) across temporal delays ($n$-back), computed with an identical trial-level pipeline for model predictions ($N=100$ models, $\sigma=0.3$), our in-house human data ($N=83$), and an independent public dataset \citep{Pascucci2019-tj} ($N=20$). Following Esposito et al.~\citep{Esposito2026-ac}, the 1-back amplitude is left uncorrected and the correction for response autocorrelation is applied at lags $\ge 2$. Asterisks denote statistical significance ($p < 0.05$, one-sided one-sample $t$-test against zero).}
\label{tab:human_model_comparison}
\resizebox{\textwidth}{!}{%
\begin{tabular}{lcccccc}
\toprule
$n$-back & Model ($^\circ$) & Model $p$ & In-house ($^\circ$) & In-house $p$ & Pascucci ($^\circ$) & Pascucci $p$ \\
\midrule
1-back & $1.807 \pm 0.018$ & $<0.001^*$ & $2.728 \pm 1.255$ & $0.016^*$ & $1.751 \pm 0.362$ & $<0.001^*$ \\
2-back & $0.408 \pm 0.020$ & $<0.001^*$ & $2.123 \pm 1.162$ & $0.036^*$ & $0.247 \pm 0.592$ & $0.340$ \\
3-back & $0.183 \pm 0.027$ & $<0.001^*$ & $0.469 \pm 0.923$ & $0.306$ & $0.163 \pm 0.595$ & $0.394$ \\
\bottomrule
\end{tabular}%
}
\end{table}

\subsection{Emergence of a spatial continuity field}
\begin{figure}[tp]
  \centering
  \includegraphics[width=0.48\textwidth]{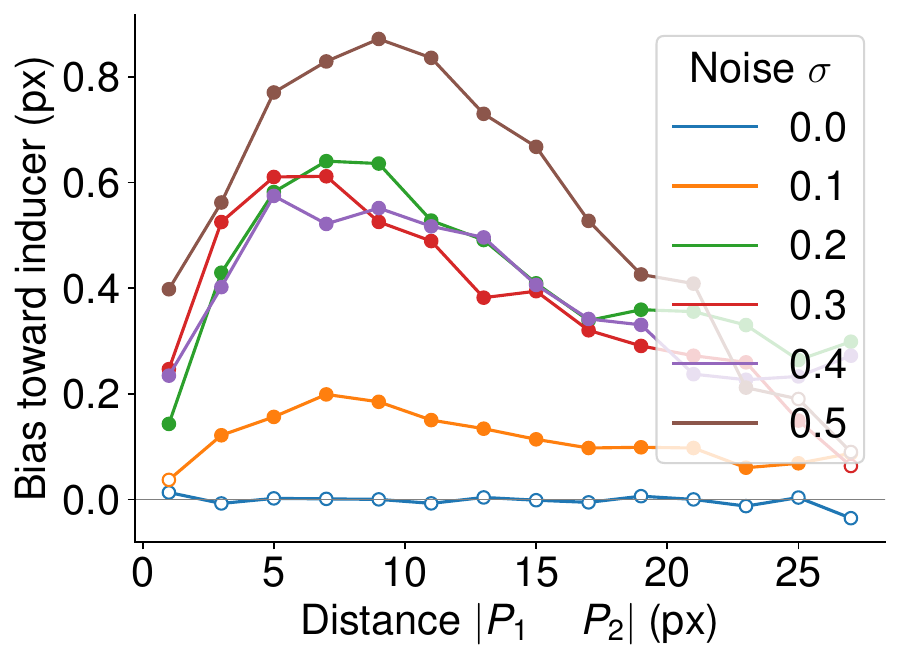}
  \caption{
    Spatial serial dependence (Continuity Field).
    Positional bias toward the inducer is plotted as a function of the spatial distance $|P_1 - P_2|$ between the inducer position $P_1$ (previous stimulus) and the target position $P_2$ (current stimulus); positive values indicate attraction.
    Filled dots mark biases significantly different from zero ($p < 0.01$, one-sample $t$-test); open dots are non-significant.
    Models trained with noise exhibit an attractive continuity field that is strongest at short distances and decays with separation, growing in magnitude with training noise, whereas the noise-free model ($\sigma=0.0$) shows no bias at any distance.
  }
  \label{fig:position_sd}
\end{figure}

Beyond orientation, we investigated whether the model develops a spatial ``continuity field'' \citep{Fischer2014-rv} by evaluating serial dependence in spatial position.
To this end, we presented two-frame sequences in which an inducer dot at one location was followed by a target dot at another, estimated the target's perceived position as the intensity-weighted centroid of the reconstruction, and measured its positional bias as the projection of the estimation error onto the direction from the target to the inducer, such that positive values indicate attraction toward the previously seen location (Materials and methods).
Models trained under noise developed a clear spatial continuity field (Figure \ref{fig:position_sd}): a significant attraction toward the recently seen location that was strongest when the two stimuli were close and declined monotonically as their separation increased.
The magnitude of this attraction grew with training noise, being absent in the noise-free model and largest at $\sigma=0.5$.
For instance, the $\sigma=0.3$ model showed a peak attractive bias of $0.61$ px at a distance of 7~px ($t(461) = 35.4, p < 0.001$, one-sample $t$-test), which declined with separation to $0.15$ px at 25~px ($t(73) = 4.17, p < 0.001$).
In contrast, the model trained without noise ($\sigma=0.0$) showed no significant spatial bias at any distance (e.g., $0.001$ px at 7~px, $t(461) = 0.20, p = 0.84$).

\begin{figure}[tp]
    \centering
    \begin{minipage}{0.48\textwidth}
        \centering
        \includegraphics[width=\textwidth]{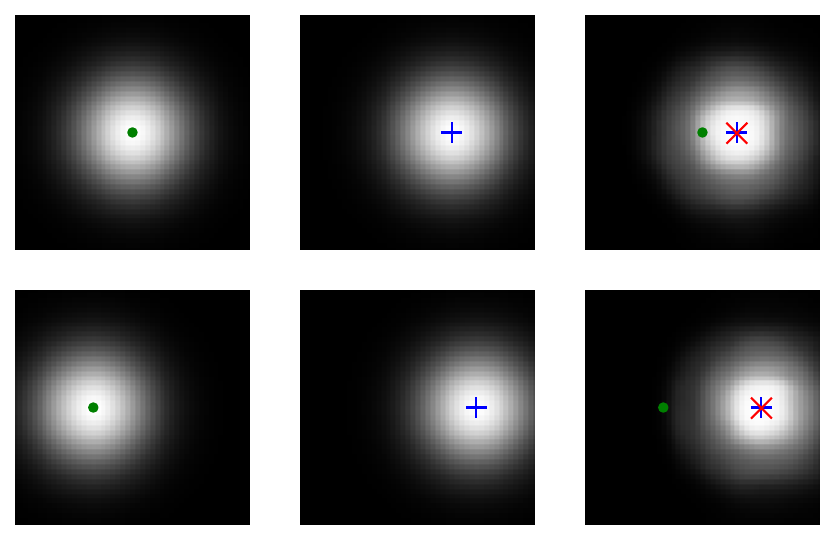}
        \subcaption{}
        \label{fig:flow_noise0}
    \end{minipage}
    \hfill
    \begin{minipage}{0.48\textwidth}
        \centering
        \includegraphics[width=\textwidth]{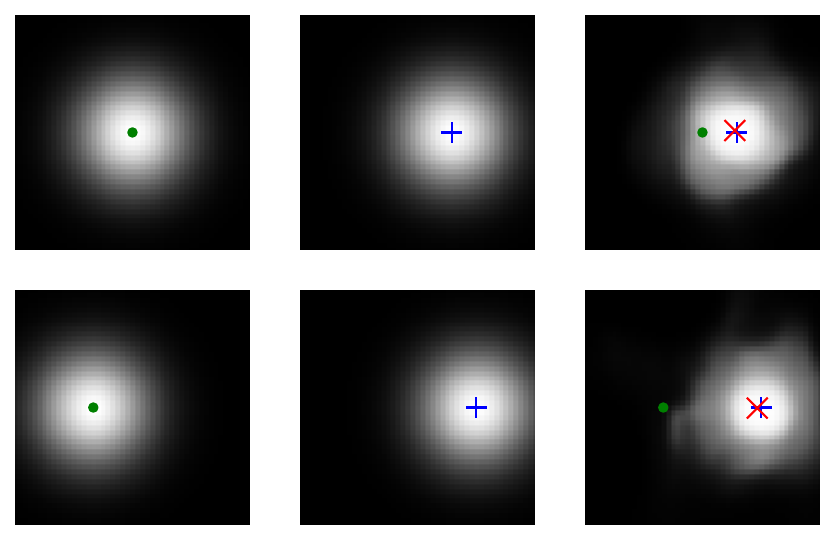}
        \subcaption{}
        \label{fig:flow_noise4}
    \end{minipage}
    \caption{
      Visualization of spatial serial dependence.
      Columns from left to right: inducer, target, and reconstructed target with the true target center (blue $+$), the estimated centroid (red $\times$), and the inducer location (green dot).
      The top row shows a near arrangement of the inducer and target, and the bottom row a far arrangement.
      (a) The model trained without noise ($\sigma=0.0$) shows no bias toward the inducer in either arrangement.
      (b) The model trained under noise ($\sigma=0.3$) exhibits a robust attractive bias toward the inducer when the two stimuli are close (top), whereas the bias diminishes when they are far apart (bottom).
    }
    \label{fig:vis_pos_flow}
\end{figure}

Figure \ref{fig:vis_pos_flow} visualizes these distance-dependent biases: target reconstructions are pulled toward nearby inducers, and the pull fades as the inducer recedes.
These results suggest that spatial serial dependence functions as an adaptive strategy: the model learns to integrate spatially proximal information, which is most likely to originate from the same underlying source, while discounting distant, likely irrelevant, context.

\section{Discussion}
Our findings support the hypothesis that serial dependence is a functional byproduct of achieving noise robustness.
Whereas prior mechanistic work showed \textit{how} a residual memory trace can produce serial dependence at the algorithmic level \citep{Higashi2025-sd}, our results supply the complementary computational-theory account of \textit{why} such traces are maintained \citep{Marr1982}: they implement an adaptive denoising strategy whose value grows with sensory uncertainty.
Framing the phenomenon normatively also clarifies that serial dependence is a property of the \textit{objective} (stable reconstruction under noise) rather than of any particular circuit, and should therefore recur across otherwise dissimilar systems that share this objective.

The force of this account rests on what temporal memory alone does not entail.
Any system with a recurrent state can carry information forward, but that does not fix the \textit{form} of the resulting bias.
A reconstruction objective is not guaranteed to yield an orientation bias tuned to small relative differences, a spatial pooling that declines with distance, or a temporal decay whose magnitude tracks human observers.
That these independently measured signatures emerge together from optimizing for noise robustness, and quantitatively match human data the model was never fit to, is what distinguishes a functional explanation from a redescription of the mechanism.

Mechanistic models of serial dependence typically fall into two families: recurrent-dynamics models, in which the bias arises from persistent neural activity or activity-silent synaptic traces within working memory \citep{Barbosa2020-tp, Stokes2015-am}, and readout or two-stage models, in which early sensory areas adapt repulsively while attractive serial dependence is introduced at a higher-level readout that integrates current input with a temporal prior \citep{Fritsche2017-yn, Pascucci2019-tj}.
Our recurrent autoencoder bridges these perspectives: the ConvGRU module provides the persistent state that maintains stimulus history, while end-to-end optimization shapes the readout mapping this state onto the reconstructed percept.
Functionally, this readout realizes the reliability-weighted integration formalized by continuity-field \citep{Manassi2024-ob} and Kalman-filter \citep{Cicchini2024-jp} accounts, in which the optimal estimate is a reliability-weighted average of current input and the state predicted from previous context \citep{Meyniel2017-jz, Ernst2002-hu}.
As training noise $\sigma$ rises, the network places greater weight on this internal prior, behaving as a nonlinear optimal filter whose gain is set by input uncertainty, and thereby produces stronger, human-like attractive biases \citep{Stocker2006-nz}.

This optimization also yielded a spatially tuned continuity field \citep{Fischer2014-rv}.
Contextual integration was strongest for nearby stimuli and fell off with distance, so that the model pooled information selectively when spatial proximity suggested that the two stimuli belonged to the same physical source.
Such a trade-off between stability and accuracy is characteristic of Bayesian inference in human vision \citep{Cicchini2024-jp}: by smoothing out momentary fluctuations, the model maintains a coherent scene representation at the cost of absolute veridicality \citep{Fischer2014-rv, Pascucci2019-tj}.
Unlike the repulsive surround reported in some human spatial-position studies, our reconstruction-only objective produced a purely attractive field that decayed with distance rather than reversing sign, an asymmetry we return to below.

Framed as a normative principle, this account generates concrete, testable predictions for biological vision.
First, because the emergent bias tracks the level of training noise, it predicts that human serial dependence should intensify under greater sensory uncertainty (for instance at low contrast, brief presentation, or high external noise), consistent with the reliability-weighting logic of optimal-integration accounts \citep{Cicchini2024-jp, Stocker2006-nz}.
Second, because the same objective produced a distance-tuned spatial continuity field, it predicts that the spatial reach of integration should reflect the correlation structure of natural scenes, pooling proximal signals likely to share a common source while discounting distant ones, rather than having a fixed extent.
These predictions tie the functional objective directly to measurable perceptual (and, in principle, neural) signatures, and remain agnostic to the specific implementing substrate.

Our direct comparison with human behavior provides quantitative support for this functional hypothesis, and an initial apparent mismatch proved instructive.
The raw in-house human profile was non-monotonic, with an anomalous 2-back peak absent from the model.
This peak, however, largely reflected a well-documented statistical confound, autocorrelation among successive adjustment responses, rather than a genuine perceptual dependence on older stimuli \citep{Esposito2026-ac}.
Once this confound was controlled, the human profile became monotonic and 1-back-dominant, matching the model; the agreement was cleanest in an independent public dataset \citep{Pascucci2019-tj} (albeit a modest sample, $N=20$), whose corrected 1-back amplitude was statistically indistinguishable from the model's (Figure \ref{fig:model_vs_human}).
That a network optimized solely for noise-robust reconstruction reproduces the corrected human profile on data it was never fit to is strong evidence that attractive serial dependence is a functional signature of this objective rather than an idiosyncrasy of either system.
A residual 2-back attraction did persist in our in-house data after correction; because single-step response correction is conservative, we regard the 2-back peak as largely, rather than wholly, an artifact and leave the origin of this residual open.

Several limitations remain.
First, although we used independent Gaussian noise, real-world visual noise is highly structured (e.g., motion blur, occlusions). Whether structured noise induces similar continuity fields is an important future direction.
Second, reproducing these phenomena in biologically plausible networks with local synaptic learning rules would further substantiate our account.
Third, human vision operates continuously across saccadic eye movements. Our model processes egocentric videos but lacks an active gaze-shift mechanism, leaving the coordinate frame (retinotopic vs.\ spatiotopic) of the emergent biases an open question.
Finally, human perception exhibits a complex interplay between attractive serial dependence and repulsive adaptation. Because our model was optimized solely for reconstruction, it developed purely attractive biases: its spatial continuity field decayed toward zero with distance but did not reverse into the repulsive surround seen in some human data, and its orientation biases showed no repulsive component. Future models should incorporate diverse behavioral objectives, such as change detection or discrimination, to capture the dynamic balance between sensory tracking and stabilization that gives rise to repulsion.

In summary, this study shows that serial dependence can emerge as a natural consequence of optimizing for robust reconstruction under sensory noise.
By supplying a normative, computational-theory account that complements existing mechanistic models \citep{Higashi2025-sd}, it reframes perceptual history-dependence not as a bias to be explained away but as a signature of adaptive temporal inference, and advances the use of task-optimized models to uncover the functional principles of biological vision.

\section{Materials and methods}

\subsection{Model architecture}
The recurrent autoencoder is designed to process video sequences, as illustrated in the top panel of Figure \ref{fig:architecture}.
The model operates on $64 \times 64$ grayscale frames and consists of three primary components:
\begin{itemize}
    \item \textbf{Encoder}: A CNN that compresses the input into a $128 \times 8 \times 8$ feature map.
    It begins with a $3 \times 3$ convolutional layer (32 channels) followed by Batch Normalization and ReLU, then three Residual Blocks with stride 2 (each block: two $3 \times 3$ convolutions with Batch Normalization), progressively increasing the channel depth to 64, 128, and finally 128.
    \item \textbf{Memory (ConvGRU)}: A Convolutional Gated Recurrent Unit with a $3 \times 3$ kernel and 128 hidden channels.
    It maintains a temporal state $h_t$, integrating the current encoder features with the state from the previous frame.
    \item \textbf{Decoder}: A CNN that reconstructs the clean image from the updated latent state.
    It consists of three Residual Blocks paired with bilinear upsampling layers, progressively decreasing the channel depth (128, 64, 32) while restoring the spatial resolution to $64 \times 64$, followed by a $3 \times 3$ convolutional layer and a sigmoid activation to produce the final output in $[0, 1]$.
\end{itemize}

Formally, let $x^{(\mathrm{noisy})}_t \in \mathbb{R}^{C \times H \times W}$ be the noisy input frame at time step $t$.
The encoder $E$ extracts the latent feature representation $z_t = E(x^{(\mathrm{noisy})}_t)$.
The ConvGRU memory module updates its hidden state $h_t$ to integrate the temporal context. The update equations are
\begin{align}
    u_t &= \sigma(W_u \ast [z_t, h_{t-1}]) \\
    r_t &= \sigma(W_r \ast [z_t, h_{t-1}]) \\
    \tilde{h}_t &= \tanh(W_h \ast [z_t, r_t \odot h_{t-1}]) \\
    h_t &= (1 - u_t) \odot h_{t-1} + u_t \odot \tilde{h}_t
\end{align}
where $\ast$ denotes a convolution operation, $[\cdot, \cdot]$ denotes concatenation along the channel dimension, $\odot$ is the Hadamard product, and $\sigma$ is the sigmoid activation function.
The current feature $z_t$ and the updated memory state $h_t$ are then concatenated and fused via a $1 \times 1$ convolution $W_{\mathrm{combine}}$ to form the context-aware latent state $z'_t$:
\begin{equation}
    z'_t = W_{\mathrm{combine}} \ast [z_t, h_t]
\end{equation}
Finally, the decoder $D$ reconstructs the clean image $\hat{x}_t$ from this updated latent state:
\begin{equation}
    \hat{x}_t = D(z'_t)
\end{equation}

\subsection{Training procedure}
The models were trained on the first-person (Ego) videos from the Charades-Ego dataset \citep{Sigurdsson2018-charadesego}.
During training, each clean video frame $x^{(\mathrm{clean})}_t$ was corrupted by additive Gaussian noise to generate the noisy input:
\begin{equation}
  x^{(\mathrm{noisy})}_t = \mathrm{clip}\!\left(x^{(\mathrm{clean})}_t + \epsilon_t,\; 0,\; 1\right), \quad \epsilon_t \sim \mathcal{N}(0, \sigma^2)
\end{equation}
The network was trained to minimize the Mean Squared Error (MSE) between the reconstructed frames $\hat{x}_t$ and the original clean frames $x^{(\mathrm{clean})}_t$ over a sequence of length $T=16$:
\begin{equation}
    \mathcal{L} = \frac{1}{T} \sum_{t=1}^T \left\| \hat{x}_t - x^{(\mathrm{clean})}_t \right\|_2^2
\end{equation}
The optimization was performed for 100 epochs using the Adam optimizer \citep{Kingma2014-ng} with a learning rate of $1 \times 10^{-3}$ and a batch size of 32.
We trained six separate models, systematically varying the standard deviation of the training noise $\sigma \in \{0.0, 0.1, 0.2, 0.3, 0.4, 0.5\}$.
All models were implemented in PyTorch \citep{Paszke2019-pj}.

\subsection{Evaluation of serial dependence}
To evaluate the emergence of serial dependence, we measured biases in both orientation and spatial position.

\paragraph{Orientation serial dependence.}
To measure serial dependence in orientation, we froze the trained weights and presented 500 independent sequences of Gabor stimuli to each model (Figure \ref{fig:architecture}, bottom; 200 per model for the 100-model robustness analysis described below, matching the 200 trials completed by each human participant).
Each sequence consisted of 5 frames of Gabor stimuli ($64 \times 64$ pixels, $\sigma = 10, \lambda = 10, \psi = 0, \gamma = 1$; orientations drawn uniformly from $[0^\circ, 180^\circ)$).
The perceived orientation of the last frame was estimated by template matching: the reconstructed image was correlated (Pearson $r$) against pre-computed Gabor templates at integer degrees from $0^\circ$ to $179^\circ$, and the orientation with the highest correlation was taken as the estimate.
The adjustment error (bias) was defined as the signed difference between the estimated and true orientation of the last frame, wrapped to $(-90^\circ, 90^\circ]$.
Serial dependence strength was quantified by fitting a Derivative of Gaussian (DoG) function:
\begin{equation}
    y = \sqrt{2e} \cdot a \cdot w \cdot x \cdot e^{-(wx)^2}
\end{equation}
where $a$ is the peak amplitude (used as the primary metric for SD strength), $w$ is a width parameter, and $x$ is the relative orientation between the current and preceding stimulus.
The statistical significance of each model's SD amplitude was assessed by a permutation test (1000 permutations): the labels of relative angles were shuffled and the DoG amplitude was re-fitted; the reported $p$-value is the proportion of permuted amplitudes exceeding the observed amplitude.
For the group-level comparison of $n$-back amplitudes (across the 100 models and across human participants; Table \ref{tab:human_model_comparison}), significance was instead assessed by a one-sided, one-sample $t$-test of the per-model (or per-participant) DoG amplitudes against zero.

\paragraph{Controlling for response autocorrelation in the $n$-back analysis.}
Serial dependence at lags $\ge 2$ can be inflated by autocorrelation among successive responses: because adjacent responses are themselves correlated, an apparent dependence of the current response on the 2- or 3-back \textit{stimulus} may be inherited from its dependence on the 1-back \textit{response} \citep{Esposito2026-ac}.
To control for this, we computed $n$-back amplitudes following Esposito et al.~\citep{Esposito2026-ac}: for each lag $k \ge 2$ we regressed out the effect of the 1-back response and refitted the DoG to the residual dependence on the $k$-back stimulus, whereas the 1-back amplitude (collinear with the 1-back response, and hence over-corrected by this procedure) was left uncorrected.
We refer to the resulting profile (1-back uncorrected; lags $\ge 2$ corrected) as the ``corrected'' profile and use it throughout for all $n$-back comparisons.
To place the model on the same footing as human observers, we treated each sequence's final-frame reconstruction as the ``current response'' and the preceding-frame reconstruction as the ``previous response'', then passed the resulting trial table through the identical human analysis pipeline; all amplitudes were computed at the level of individual trials (trial-level DoG).

\paragraph{Human behavioral data.}
To independently validate the model predictions against human behavior, we analyzed a dataset from an orientation adjustment experiment first reported in our previous study~\citep{Higashi2025-sd}.
The study was approved by the Committee for Human Research at the Graduate School of Engineering, The University of Osaka (approval number 5-4-1), and conducted in accordance with the Declaration of Helsinki.
One hundred participants were recruited via Prolific (\url{www.prolific.com}) and provided informed consent prior to participation; they received a base compensation of \pounds3 plus a performance-based bonus of up to \pounds1.

Stimuli were sinusoidal gratings generated using PsychoPy \citep{Peirce2019}, presented centrally with a Gaussian aperture.
On each trial, spatial frequency was drawn uniformly from $\{2, 3, 4\}$ cycles/degree, orientation from $\{0^\circ, 1^\circ, \ldots, 179^\circ\}$, and phase from $\{0.0, 0.1, \ldots, 0.9\}$.
Each grating was displayed for either 0.3~s or 0.6~s, immediately followed by a 0.5-s mask to suppress aftereffects.
Participants adjusted a keyboard-controlled bar (rotated via the left and right cursor keys and confirmed with the space bar) to match the orientation of the just-seen grating; the bar's initial orientation was randomized.
Each participant completed 200 trials arranged in four blocks of 50, with a mandatory rest of at least 30~s between blocks (mean session duration: 27~min).

Participants whose mean absolute orientation error exceeded $39.2^\circ$ ($1.5 \times \mathrm{IQR}$ above the third quartile) were excluded, removing 17 of 100 participants (17\%).
Trials with an absolute error above $37.5^\circ$ (same criterion applied trial-wise) were additionally removed (488 trials, $\approx$2.9\%), yielding a final dataset of 16,112 trials from $N = 83$ participants.
For the present analysis, adjustment errors were pooled across spatial frequency and presentation-time conditions, and $n$-back serial dependence ($n=1,2,3$) was computed at the trial level with the response-autocorrelation-corrected pipeline described above, identically to the model.

\paragraph{Independent public dataset.}
As an independent test, we reanalyzed the publicly available orientation-adjustment data of Pascucci et al.~\citep{Pascucci2019-tj} (Zenodo, doi:10.5281/zenodo.2544946).
We used the two baseline experiments without additional within-session condition manipulations, Experiment~1 (10 observers, discrete $18$-orientation sampling) and Experiment~2 (10 observers, continuous orientations), pooled into a single sample of $N=20$ observers to maximize statistical power.
Raw responses were converted to the same signed-error and relative-orientation conventions as our in-house data and passed through the identical trial-level, response-autocorrelation-corrected pipeline.

\paragraph{Spatial serial dependence.}
To evaluate spatial serial dependence (often termed a ``continuity field''), we used Gaussian dot stimuli ($\sigma=8.0$ px).
We generated 5,000 two-frame sequences: the first frame contained the inducer dot at position $P_1$ and the second contained the target dot at position $P_2$, each placed at a random location drawn uniformly from the range $[20, 44]$ px in both dimensions (to avoid edge effects).
The model was fed each sequence, and the perceived spatial position of the target was estimated from the reconstructed second frame using centroid estimation (center-of-mass weighted by pixel intensities above 10\% of the image maximum).
The positional bias was defined as the projection of the estimation error vector (estimated minus true target position) onto the unit vector pointing from the target to the inducer, such that positive values indicate attraction toward the inducer.
Bias was analyzed as a function of the Euclidean distance $|P_1 - P_2|$ between the inducer and target positions.

\section*{Acknowledgments}
This work was supported in part by the Japan Society for the Promotion of Science (JSPS) KAKENHI, grant numbers 22H05163 and 24K15047,
and Japan Science and Technology Agency (JST) Advanced International Collaborative Research Program (AdCORP), grant number JPMJKB2307.

\section*{Author contributions}
H.H. and K.S. conceived and designed the study. H.H. implemented the model and performed all simulations and analyses. H.H., K.S., and Y.T. wrote the manuscript.

\section*{Data and code availability}
All code and data supporting this study are available at \url{https://github.com/labbip/noise_sd}. The independent human dataset reanalyzed here is publicly available from Pascucci et al.~\citep{Pascucci2019-tj} (Zenodo, doi:10.5281/zenodo.2544946).

\section*{Competing interests}
The authors declare no competing interests.

\bibliographystyle{plos2015}
\bibliography{references}

\begin{thebibliography}{10}

\bibitem{Fischer2014-rv}
Fischer J, Whitney D.
\newblock Serial dependence in visual perception.
\newblock Nature Neuroscience. 2014;17(5):738--743.

\bibitem{Burr2014-uo}
Burr D, Cicchini GM.
\newblock Vision: Efficient adaptive coding.
\newblock Current Biology. 2014;24(22):R1096--R1098.

\bibitem{Fritsche2017-yn}
Fritsche M, Mostert P, de~Lange FP.
\newblock Opposite effects of recent history on perception and decision.
\newblock Current Biology. 2017;27(4):590--595.

\bibitem{Murai2021-sd}
Murai Y, Whitney D.
\newblock Serial dependence revealed in history-dependent perceptual templates.
\newblock Current Biology. 2021;31(14):3185--3191.

\bibitem{Alais2017-tn}
Alais D, Leung J, Van~der Burg E.
\newblock Linear summation of repulsive and attractive serial dependencies:
  Orientation and motion dependencies sum in motion perception.
\newblock Journal of Neuroscience. 2017;37(16):4381--4390.

\bibitem{Manassi2018-be}
Manassi M, Liberman A, Kosovicheva A, Zhang K, Whitney D.
\newblock Serial dependence in position occurs at the time of perception.
\newblock Psychonomic Bulletin \& Review. 2018;25(6):2245--2253.

\bibitem{Cicchini2014-sg}
Cicchini GM, Anobile G, Burr DC.
\newblock Compressive mapping of number to space reflects dynamic encoding
  mechanisms, not static logarithmic transform.
\newblock Proc Natl Acad Sci U S A. 2014;111(21):7867--7872.

\bibitem{Hashiguchi2026}
Hashiguchi K, Tanaka Y, Higashi H.
\newblock Sequential biases in the perception of proportion.
\newblock Journal of Vision. 2026;26(11):7.

\bibitem{Liberman2014-kp}
Liberman A, Fischer J, Whitney D.
\newblock Serial dependence in the perception of faces.
\newblock Current Biology. 2014;24(21):2569--2574.

\bibitem{Liberman2018-tb}
Liberman A, Manassi M, Whitney D.
\newblock Serial dependence promotes the stability of perceived emotional
  expression depending on face similarity.
\newblock Attention, Perception, \& Psychophysics. 2018;80(6):1461--1473.

\bibitem{Higashi2025-sd}
Higashi H.
\newblock A residual memory trace in an accumulator explains serial dependence.
\newblock bioRxiv. 2025;2025.08.24.671986.

\bibitem{Marr1982}
Marr D.
\newblock Vision: A Computational Investigation into the Human Representation
  and Processing of Visual Information.
\newblock San Francisco: W. H. Freeman; 1982.

\bibitem{Manassi2024-ob}
Manassi M, Whitney D.
\newblock Continuity fields enhance visual perception through positive serial
  dependence.
\newblock Trends in Cognitive Sciences. 2024;28(6):497--511.

\bibitem{Cicchini2024-jp}
Cicchini GM, Mikellidou K, Burr DC.
\newblock Serial Dependence in Perception.
\newblock Annual Review of Psychology. 2024;75:231--255.

\bibitem{Stocker2006-nz}
Stocker AA, Simoncelli EP.
\newblock Noise characteristics and prior expectations in human visual speed
  perception.
\newblock Nature Neuroscience. 2006;9(4):578--585.

\bibitem{Ernst2002-hu}
Ernst MO, Banks MS.
\newblock Humans integrate visual and haptic information in a statistically
  optimal fashion.
\newblock Nature. 2002;415(6870):429--433.

\bibitem{Kriegeskorte2018-ir}
Kriegeskorte N, Douglas PK.
\newblock Cognitive computational neuroscience.
\newblock Nat Neurosci. 2018;21(9):1148--1160.

\bibitem{Tang2018-uj}
Tang H, Schrimpf M, Lotter W, Moerman C, Paredes A, Ortega~Caro J, et~al.
\newblock Recurrent computations for visual pattern completion.
\newblock Proc Natl Acad Sci U S A. 2018;115(35):8835--8840.

\bibitem{Esposito2026-ac}
Esposito D, Fornaciai M, Gori M.
\newblock Investigating the effect of response autocorrelation on n-back
  analyses of serial dependence.
\newblock Journal of Vision. 2026;26(1):12.

\bibitem{Pascucci2019-tj}
Pascucci D, Mancuso G, Santandrea E, Della~Libera C, Plomp G, Chelazzi L.
\newblock Laws of concatenated perception: Vision goes for novelty, decisions
  for perseverance.
\newblock PLOS Biology. 2019;17(3):e3000144.

\bibitem{Barbosa2020-tp}
Barbosa J, Stein H, Martinez RL, Galan-Gadea A, Li S, Dalmau J, et~al.
\newblock Interplay between persistent activity and activity-silent dynamics in
  the prefrontal cortex.
\newblock Science. 2020;369(6505):eaba7699.

\bibitem{Stokes2015-am}
Stokes MG.
\newblock {`Activity-silent'} working memory in prefrontal cortex: a dynamic
  coding framework.
\newblock Trends in Cognitive Sciences. 2015;19(7):394--405.

\bibitem{Meyniel2017-jz}
Meyniel F, Dehaene S.
\newblock Brain networks for confidence weighting and hierarchical inference
  during probabilistic learning.
\newblock Proc Natl Acad Sci U S A. 2017;114(19):E3859--E3868.

\bibitem{Sigurdsson2018-charadesego}
Sigurdsson GA, Gupta A, Schmid C, Farhadi A, Alahari K.
\newblock Actor and Observer: Joint Modeling of First and Third-Person Videos.
\newblock In: Proceedings of the IEEE Conference on Computer Vision and Pattern
  Recognition (CVPR); 2018. p. 7396--7404.

\bibitem{Kingma2014-ng}
Kingma DP, Ba J.
\newblock Adam: A method for stochastic optimization.
\newblock arXiv [csLG]. 2014;.

\bibitem{Paszke2019-pj}
Paszke A, et~al.
\newblock {PyTorch}: An imperative style, high-performance deep learning
  library.
\newblock arXiv [csLG]. 2019;.

\bibitem{Peirce2019}
Peirce J, Gray JR, Simpson S, MacAskill M, H{\"{o}}chenberger R, Sogo H, et~al.
\newblock {PsychoPy2: Experiments in behavior made easy}.
\newblock Behavior Research Methods. 2019;51(1):195--203.

\end{thebibliography}

\end{document}